\documentclass[10pt, a4paper]{article}

\def\normalbaselineskipp{5.05mm plus.11mm}

\def\smallbaselineskipp{4.4mm}

\def\usenormalbaselineskipz{\baselineskip=\normalbaselineskipp}

\def\usesmallbaselineskipz{\baselineskip=\smallbaselineskipp}

\font\titletype=cmr10 at 15pt

\font\authortype=cmr10

\font\headingboldtype=cmb10 at 12pt

\font\normaltype=cmr10

\font\bf=cmb10

\font\it=cmti10

\def\rulez{\noindent\hrule width \hsize height.2mm}

\def\tocsectionitemz#1#2{\noindent\rlap{\bf #1}\kern 5mm{\it #2}}

\def\numberedsectionz#1#2{\vskip 12mm \centerline{\headingboldtype #1\kern 3.5mm #2} \nobreak \vskip 3.3mm}

\def \referencesz{\vskip 13mm \centerline{\headingboldtype
References} \nobreak \vskip 2.2mm \usesmallbaselineskipz}

\begin{document}

\thispagestyle{empty}

\usenormalbaselineskipz

\hrule height 0pt

\vskip -5mm

\vskip7mm\centerline{\titletype The Hourglass---Consequences of Pure}

\vskip -4mm

\vskip7mm\centerline{\titletype Hamiltonian Evolution of a Radiating System}

\vskip 3mm

\centerline{\authortype Donald McCartor}

\vskip 10.5mm

\rulez 

\vskip 6mm

\normaltype

\usesmallbaselineskipz

\centerline{ABSTRACT}

\vskip 1.35mm

\noindent Hourglass is the name given here to a formal isolated quantum system that can radiate.  Starting from a time when it defines the system it represents clearly and no radiation is present, it is given straightforward Hamiltonian evolution.  The question of what significance hourglasses have is raised, and this question is proposed to be more consequential than the measurement problem.

\vskip 2.3mm

\tocsectionitemz{1}{Hourglasses}

\tocsectionitemz{2}{Physics without true histories}

\tocsectionitemz{3}{But histories are sometimes good}

\tocsectionitemz{4}{Phlogiston and oxygen}

\tocsectionitemz{5}{A closer look at quantum engineering}

\tocsectionitemz{6}{Conclusion}

\vskip 3.6mm

\rulez

\normaltype

\usenormalbaselineskipz

\vskip 10mm

\hfill\parbox{8cm}{But I want to know the {\it particular\/} go of it}

\vskip 3mm

\hfill\parbox{8cm}{\sloppy -- the plea of James Clerk Maxwell as a young child concerning, among many things, the bell-wires that ring the bells that summon servants. [Mahon]}

\vskip-5mm

\numberedsectionz{1}{Hourglasses}

\noindent Suppose that theory develops in such a way that quantum fields can be handled like nonrelativistic quantum mechanics.  Then if we are interested in something, perhaps gooseberry bushes, we can model one as we would conceive it to be at some instant and then follow its development through time.  And not only the atoms and molecules would be modeled, but also the radiation.

This is a scheme for the imagination.  The gooseberry bush, though not isolated, would grow within a suitable environment that would be an isolated system, complete in itself.  We do learn well from isolated systems, both real ones in the laboratory and those envisaged in our theoretical musings.

We will provide the bush with air, earth, and water.  And there can be life-giving sunlight shining on it.  As for the light that had been reflected or emitted from the bush before the present time, we will leave that out.  Such light goes off and away, so it could only matter as information about what the bush had been doing.  We will take the bush just as it is now.

The gooseberry bush is then developed forward in time.  Lagrange or Hamilton would have recognized what we are doing, for we are doing physics the classical way.  We have an initial condition and we are finding out what will happen next.

As we move toward the future, light shoots out from the bush, as we expect.  But, disconcertingly, the bush starts to lose definition.  Its parts lose their precise places.  Within a few weeks it is a scarcely recognizable mess.  Let\rq s go back in time, then.  This is terrifyingly worse.  The bush has been the subject of a vast conspiracy.  Light has been streaming in on it from the entire universe.  The bush swallows it up.  Then at the present time this suddenly all stops.  Time symmetry of the Hamiltonian makes it happen like that.

This is the hourglass.  It is really more like a cone, with the light streaming in before the set-up time forming one nappe and the light streaming out after it the other.  But hourglass is a more colorful name.

What to do? We will try to bring quantum mechanics to the rescue.  We will make what is conventionally called a measurement, but cautiously.  A place is chosen well outside the gooseberry bush, and a time chosen that is later than when we set the state of the bush up.  A check is made of whether there are at this place and time any photons coming from the direction of the bush.  By doing things this way, we won\rq t disturb the bush at all, and we don\rq t care if we disturb the escaping light.  We get from this, of course, a probability distribution over various possibilities for photons at this place and time.  Encouraged by this small success, we choose another place and time and do the same.  And this is what is nice: the two measurements are compatible.  Thus we get correlations between them, too.  Emboldened by this opportunity, we do millions of them, which all formally combine into a single measurement with a single set of possible results.  Each possible result of the single, combined measurement is a combination of  results of all the individual measurements of light made at the various times and places.  Thus each combined result constitutes a kind of movie of the gooseberry bush.

What will the most probable of these results be like?  This is the problem of the hourglass.  To begin with, however, it may be that there is no hourglass.  The deepest quantum theory might not provide a system with a state and its evolution.  Or if it does, it could still be objected that the Hamiltonian evolution should not have been allowed to run on unchecked.  There should have been many quantum jumps.  By leaving them out, quantum mechanics has been misused, and what results is no matter.

But Lagrange and Hamilton and would have been best pleased if these objections did not hold.  And surely we would then hope to see in each of the most probable results something like a movie of a bush producing gooseberries: physics working right.  The bushes in these movies would look much alike at the start but then gradually differ, as chance has it.  We would learn something about how gooseberry bushes grow gooseberries!

Certainly Lagrange and Hamilton would have thought the problem of the hourglass a leading one, if they had known of quantum mechanics.  Indeed, every physicist might like to take a stab at guessing its solution, just to orient themselves in their science.  Does the hourglass fail, and if so, where and why?  Or if it does produce movies true to our world, but not from a developing quantum state that might be the true history of a gooseberry bush, rather from a \lq\lq history\rq\rq{} that does at one time represent a gooseberry bush well, but soon is unlike anything that ever did exist, then how can this be?

\numberedsectionz{2}{Physics without true histories}

\noindent Here is what I think about it.  But before we go into that, see if you don\rq t agree that the hourglass question has gravity, and this regardless of the ideas that I or anyone might have for its answer.

Now my guess is that quantum mechanics will give us movies of ripening gooseberries, produced by hourglasses through the means described or something rather like that.  And I think that to understand hourglasses, not to solve the measurement problem, is the central question for the understanding of quantum mechanics.

For the measurement problem begs a question, which makes it futile.  It assumes that we learn from physics simply because physics describes well those things that exist.  Like this example from classical physics.  There exist in a gas a multitude of zipping molecules.  At any given moment, each particle has its particular position and momentum, and over time this forms their history. Physics has told us what a gas is---precisely what exists there.  This is what lets us learn about gases.  Undoubtedly this is how Boltzmann saw it.

But when we look at the statistical mechanics he produced, and even more at that of Gibbs, a person will acquire deep qualms about this viewpoint.  Boltzmann\rq s analysis of the collision of molecules seems like straightforward common sense.  He is looking at what they are likely to do.  But when Loschmidt\rq s reversibility objection is brought forward, the lucidity vanishes.

Gibbs\rq s more abstract statistical mechanics made the problem even starker.  Gibbs found beautiful mathematical form in Boltzmann\rq s (and Maxwell\rq s) work, which he generalized.  He held that thermodynamic systems should be represented as being in states that have the form of certain probability distributions over classical states.  Gibbs could not well understand what these probabilities were about, but he saw that his theory was good nevertheless.  To keep this lack of clear comprehension from poisoning work with the theory, he devised a work-around.  The axioms of probability theory are reflected in the axioms of finite set theory.  One can effectively solve problems of probability by thinking about finite sets.  So Gibbs suggested that we simply think about these probabilities in terms of sets.  The word he used was ensembles.

Gibbs described his intent in these words:  \lq\lq The application of this principle is not limited to cases in which there is a formal and explicit reference to an ensemble of systems.  Yet the conception of such an ensemble may serve to give precision to notions of probability.  It is in fact customary in the discussion of probabilities to describe anything which is imperfectly known as something taken at random from a great number of things which are completely described.\rq\rq{}[Gibbs]  

But physicists have never been able to accept gracefully that they don\rq t understand the elements of their science.  So they have been moved to think that they do understand Gibbs\rq s probabilities somehow, and this has led to two missteps.

One has been to regard the probabilities in Gibbs\rq s theory as being the result of our ignorance of the detailed state of the system we are considering.  But when a probability distribution is useful, this is a very great step up in order from chaos.  Ignorance cannot create order.  If water always boils at the same temperature, it is not our fault.  Rather than being so explained, for it is not, Gibbs\rq s theory shows that there is something deeply wrong with classical mechanics.   Classical statistical mechanics is not really a form of classical mechanics.  It is quantum mechanics being born.

 The following words of Gibbs seem to show that Gibbs himself took the view just scotched.  \lq\lq The states of the bodies which we handle are certainly not known to us exactly.  What we know about a body can generally be described most accurately and most simply by saying that it is one taken at random from a great number (ensemble) of bodies which are completely described.\rq\rq{}[Gibbs] \hspace{0.5mm} The impression that I get, though, is that Gibbs is cautiously hedging.  He is not saying plainly, as he might have, that a body we handle {\it will be\/} in some completely described state, so that if we describe it with an ensemble, the probabilities in the ensemble simply represent our partial ignorance about that state.  He does say plainly that his method seems to work.

The other misstep has come about because quantum theory is a mirror of Gibbs\rq s statistical mechanics in the sense that it is based on what are probabilities in form (in other words, sets of non-negative real numbers that add up to one) and we don\rq t know what they mean in general.  It is true that we can make good sense of them as real probabilities in various special cases.  For instance, when quantum mechanics is applied to the Stern-Gerlach experiment, to see the detector react is like seeing a coin tossed.  But in the general case no such kind of experience is directly implied by these probability forms.  There are, for example, canonical distributions in quantum mechanics too, and we don\rq t ever expect to see a detector pick a pure state out of a hot cup of coffee.

We then sometimes think about these formal probabilities in terms of ensembles, just as Gibbs did, and for the same reason.  Where the formal probabilities are highest and the members of the ensemble most numerous, there the greatest significance will lie, whatever it may be.  This is fine.  But quite often physicists say that ensembles (that is to say, Gibbs\rq s work-around) provide the means to understand quantum theory.  This is clearly wrong.

But to get back to the measurement problem.  As you well know, but for explicitness I will say it anyway, to see a problem in measurement is to suppose that quantum mechanics can describe the equipment in the lab as it exists at the start of an experiment, but when the representation is continued, the equipment becomes entangled with the microscopic systems it is examining and gets smeared.  Then quantum mechanics has stopped describing what we know exists in the lab and needs to be corrected so that it will continue to describe what exists.

But it isn\rq t so that quantum mechanics, if it is to show us some predictability in nature, must provide us directly with histories of the existence of things, as by a developing wave packet.  As evidence, I offer the hourglass.

\numberedsectionz{3}{But histories are sometimes good}

\noindent If physics does not work simply because it describes what exists, and if, rather, the way of the hourglass is right, then a corollary is that how we learn about nature necessarily becomes more indirect.  We are given such information as radiation provides about something, not directly told what exists there.  And for the purpose of inferring useful rules of nature\rq s behavior, what we deal with are imagined situations that we think typical of what we want to learn about, not faithful descriptions of actual things.  No real radiating system is like an hourglass, except momentarily near the hourglass\rq s neck.

But quantum engineering may temper the truth of that judgment just a bit.  For there is also an engineering use of quantum mechanics where, somewhat as classical mechanics does it, for a time we can use a wave packet to represent the development of an actual situation we are dealing with.  But this is rather more special, for we must take care to set things up so that this will work.  The vacuum must be excellent, etc.  Isolation is important.

A simple example of quantum engineering is an ion that alternately blinks for a spell and remains dark for a spell while sitting in an ion trap that is irradiated by lasers.  You can picture the ion well enough by thinking of Schr\"{o}dinger evolution of a wave packet with occasional quantum jumps interspersed.  You might then be tempted to think that everything can be handled effectively in the same way, at least in principle.  We have just not been clever enough to find New York City\rq s wave packet and its measurement collapses.

This is trouble.  The worst of it is that you will be led to ignore hourglasses and what they imply, since clearly hourglasses cannot represent the history of things in the same manner that you have advantageously represented the history of the blinking ion.

On the other hand, imagine that decades ago physicists had taken hourglasses to their hearts, as well I think they might have.  Then they could have been tempted to look upon representing an ion in a trap by Schr\"{o}dinger evolution of a wave packet with quantum jumps as \lq following the wrong philosophy\rq{} (by trying to represent the actual histories of things with wave packets), and might have disdained to do so.  There is a lesson here.  Don\rq t take your philosophical ideas too seriously, we\rq re not good enough for that.

I believe, though, that from hourglasses you would be able to infer that Schr\"{o}dinger evolution with jumps is a simple and effective (not perfect) way to regard a blinking ion in a trap.  The hourglasses would then be in this sense the more fundamental theory.

\numberedsectionz{4}{Phlogiston and oxygen}

\noindent But what is a quantum jump?  Here is where I think the community of physicists has been careless in the use of words, perhaps mixed with real misunderstanding.  Two principles of quantum physics have been formulated.  The first principle (promoted by Dirac and von Neumann) is that when a measurement is made on a system, an immediately following measurement will give the same result.  Therefore, right after any measurement the system must be in the eigenstate corresponding to the value found.

The second principle is that if the probabilities of the possible results of all the measurements that may be made on a system are defined, then there will be a (unique) quantum state that the system may be said to be in that will yield these probabilities.  Add to this that sometimes two measurements may be made on a system without interfering with each other.  Then when one of the two measurements has a certain result this will define a conditional probability for any result of the other measurement (simply divide the probability that both results occur by the probability that this result of the first measurement occurs).  According to the second principle, then, there will be a quantum state that yields these probabilities (for the possible results of any measurement that may be made without interfering with, or suffering interference from, a given measurement that has had a certain result).

Please notice that the argument above assumes that the set of all the measurements compatible with a given measurement effectively constitutes \lq all the measurements that may be made on a system\rq{} as needed by the second principle.

Now consider a system~$A$ in the state~$\alpha.$  It is composed of two subsystems, $B$~and~$C,$ in reduced states $\beta$~and~$\gamma$ respectively.  A measurement is made on subsystem~$B$ and it has a result.  By the first principle, there is a quantum state~$\beta^\prime$ that will yield the probabilities of the possible results of any immediately following measurement that might be made on subsystem~$B.$  And by the second principle, there is a quantum state~$\gamma^\prime$ that will yield the probabilities of the possible results of any compatible measurement made on subsystem~$C.$

For the supplanting in one\rq s considerations of~$\beta$ by~$\beta^\prime$ there is the historical name \lq collapse of the wave packet\rq\hspace{-.3mm}.  For the supplanting in one\rq s considerations of~$\alpha$ by~$\gamma^\prime$ most physicists use the same phrase (or any of its several synonyms).  It would easier to think about these things if different names were used for the two.  \lq Collapse of the wave packet\rq{} might be retained for the first and, say, \lq conditioning of the wave packet\rq{} adopted for the second.

This is all the more important because the first principle is an out and out mistake by Dirac and von Neumann, whereas the second is an inalienable part of quantum mechanics.  To those two mathematically minded, and so logically minded, people, the dignity of quantum mechanics required that there be measurements, so that quantum mechanics might be real physics.  And since quantum mechanics did not say that a system had to have, before the measurement, the value found in the measurement, the dignity of measurement required that it at least have that value afterward, or what sort of measurement was this anyway?

Tacked on to this was the fact that so distressed Schr\"{o}dinger: wave packets spread interminably.  If a developing wave packet were to represent the history of a system, which they assumed to be necessary, then the spreading had to be checked, and an occasional quantum jump such as their measurement theory presupposed might do that.

And experiment lent some support.  Above all, if an electron went splat somewhere on a screen, which they regarded as a measurement by the experimenter of the electron\rq s position, then conservation of charge suggested strongly that the electron could be found subsequently thereabout.  This was the origin of the phrase \lq collapse of the wave packet\rq.  Too, the famous Stern-Gerlach experiment allows a following measurement of spin, which will give the same result as the first if the first measurement\rq s detection has been delicate enough.

But the idea of a quickly following measurement is just not well-defined in general.  And there are cases where the principle must prove false under any reasonable definition of a following measurement.  For example, a particle might lose most of its energy in those collisions that measured its energy.  Or if the momentum of a charged particle were measured by the curvature of its path in a magnetic field, the particle might end up going in the wrong direction, although this is, to be sure, correctable.  Those events called \lq\lq measurements\rq\rq{} are what they are, and if they fall short of truly being measurements of properties, so be it!

If the first principle is an error, then that leaves us with only one principle, the second, and people might then be inclined to continue to use the traditional phrase \lq collapse of the wave packet\rq{}, but now meaning the replacements the second principle defines.  This would result in the transfer, in the course of history, of the meaning of the phrase from the first principle to the second.  I think that this would have the same unhappy effect as if Lavoisier, not wishing to burden the world with a neologism, had instead given to the word phlogiston a new sense.

\numberedsectionz{5}{A closer look at quantum engineering}

\noindent The second principle has a very different flavor from the first.  For it leads to conditional probabilities, and these lend themselves to imaginative thinking.  In this mind-set you are free to take up points of view according to what you wish to learn.  The first principle, however, leads to probabilities that are thought to be the properties of real events, such as an actual toss of a coin.  You are now in a reality mind-set.  That probability is as much a part of the coin toss as is the silver of the coin, and you must deal with it.  You have no choice.  But I don\rq t mean to say that this is an absolute difference between the two principles.  Rather, they tend to lead us into these respective modes of thought, and vice versa.  Bearing this in mind, let us look at the hourglass and quantum engineering.

First consider the hourglass that represents a gooseberry bush.  By choosing one among the more probable of the results of the course of observation of light, we will select what is in effect a likely movie of such a bush.  We can look at the movie, and the marvelous algorithms of our brains will construct an idea of a gooseberry bush and follow it through its history.  We have gotten something good out of this, and we have made no use of the conditional probabilities offered by the second principle at all.  However, if we are not limited to one movie then we can use conditional probabilities as they are normally used, to explore various interesting possibilities while taking into account how likely they are when we are supplied with certain information.

Notice that we have been thinking imaginatively.  No one would suppose that we have directly grasped the reality of a gooseberry bush in our garden in this way, particularly because real gooseberry bushes do not start to exist at a special time.

Now consider quantum engineering.  By means of careful construction of the equipment a clearly defined situation can be set up where the power of wave packets to give understanding will be enhanced.    On the other hand, here there can be significant entanglement.  The power of our minds to achieve understanding through their everyday methods will be set at nought.

Then for quantum engineering, a history formed by wave packet development with occasional saltations may be a quite good route to understanding.  We would take up this idea of what exists simply because it is good enough to help us with the job at hand.  And for this case, where we find it fit to think that we are dealing with an actual system that is an evolving wave packet, and with saltations that we regard as actual events, but in a way so different from that intended by Dirac and von Neumann, then perhaps a third term, say, \lq change of the wave packet\rq{}, would be appropriate for the saltations.

These {\it changes\/} of the wave packet would differ from {\it collapses\/} of the wave packet because, although they would be thought of as real events just as collapses have been, they would be derived from {\it conditioning\/} of wave packets, in the following manner.  When a system sends out radiation (or anything else) that will not return, in one way you can consider the system of interest to be the whole, including the radiation, and in another way you can consider it to be the reduced system that does not include the radiation.  Upon observation of the radiation you will derive from the result and from the wave packet of the whole system a wave packet for the system less the radiation, and this we have called conditioning of the wave packet.  But if before the observation your interest had been focussed on the system less the radiation, and thus on its reduced wave packet, then you will have gone from one wave packet to another wave packet for the system less the radiation.  And since you are reckoning these wave packets as being portions of the system\rq s history, this looks like a quantum jump.  This is what is meant by a change of the wave packet.  There is no need to define any such change of the wave packet precisely, of course.  No more is there need to suppose that it {\it can} be defined with precision.

\numberedsectionz{6}{Conclusion}

\noindent If hourglasses cannot be true histories, how can it be that we can learn from them?  What lets them tell us how gooseberry bushes grow, when they are only momentarily like a gooseberry bush?  I haven\rq t said a word about this yet.

First of all, there is an assumption hidden behind this puzzlement of ours.  The assumption is that we have no reason to be perplexed that we can learn from things that can be {\it true\/} histories.  For if it did not seem so perfectly natural to us that we learn from true histories, then it would not appear unnatural to learn from what clearly cannot be a true history.  But I think this assumption of ours is thoughtless, and I will try to explain why.

We make judgments about when we are better informed and when less so.  The ideas we hold true when thought to be better informed are compared with those that we held when not so well informed.  In this way, through the device of taking the ideas that we presently have most confidence in as trustworthy, we try to gather how successfully our ideas tend to stack up against reality.  It is not quite so simple, however, since we know from sad experience that the ideas we now trust may fail us.  But we have the conviction, or hope, that if such happens we can land on our feet again.  We will search for still better ideas until we find something that works.


We are apt to give to this situation a logical cast.  Namely, by positing that there is a best of all possible ideas in whose direction we are headed.  This posit can be helpful.  It can give us greater confidence in our search for better ideas.  If we guess that this best idea will have a certain form, and we guess well, it can guide our search.  But there is no necessity for this posit; all we really know is what was said above.

Another thing we like to do is to find where things are and when.  Our vision, touch, and hearing do this automatically all the time, and we often give them some conscious help, say by turning the head.  When we are a teenager it is likely to occur to us that there must be a best of all possible such ideas, a complete map of where everything is, and has been, and perhaps will be too.  A further thought may cross one\rq s mind.  Maybe this is all that our world is.  For instance, if one person likes another, this should show up in that person\rq s actions, which the map will completely define.  Maybe the liking simply is those actions.

Now I will propose some physics, the red dust theory.  According to this theory the world is made up of an exceedingly large number of very fine specks of a scarlet dust.  Because of its ruddiness, the dust is extremely beautiful, if only we could see it, but we will not be concerned with that.  The red dust theory differs from most physics in that the flight of the particles does not have to satisfy a differential equation, it is merely continuous.

The interpretation of the theory is quite simple.  Where we find things there will be a crowd of these specks, and where we find vacancy they will be much sparser.  But can our world be as this theory says?  Surely it can.  There will be among its solutions one that maps the entire history of our universe with extraordinary precision.  The collisions of galaxies, the evolution of whales, the experiments in laboratories, all will be there and rightly shown.

Now you may think that the red dust theory is hopelessly bad physics and should be ignored.  It may be hopelessly bad, but it should not be ignored.  It is a benchmark.  If another physics theory is proposed, is it better than the red dust theory, and if so, just why?  This is especially pertinent if the other theory intends, as does the red dust theory, to give a precise description of all that exists.  Bohmian quantum mechanics is an example.

But what I intend to put up against the benchmark is classical mechanics.  Everyone will agree that classical mechanics is far better than the red dust theory.  You can do things with classical mechanics; you can\rq t do anything with the red dust theory.  For instance, you can pull a pendulum to the side and let it go.  It will swing.  Classical mechanics can give you the history of that swing ahead of time.  The red dust theory has so many solutions compatible with the way things are at the start that it won\rq t tell you anything useful about how things will go.

Our experience with classical mechanics is that it is practical, but why is this so?  The most natural idea is that the world must at bottom be classical mechanical.  Since we understood the pendulum by assigning a classical mechanical state to it and evolving the state, there must then be an evolving classical mechanical state that the whole world is in, and that would explain why classical mechanics is so useful.

When we look at the history of our universe, however, and particularly at the evolution of life over billions of years, and when we consider the resources that it is likely that classical mechanics has to offer in its solutions, it doesn\rq t really seem possible that there is any classical mechanical history that would match our universe\rq s history, no matter how exquisitely the initial conditions are chosen.  For the more detailed structures of the classical representation must in time dissolve into lasting chaos, and I would think rather quickly.

Still, this does depend on a point I don\rq  t actually know the answer to. For in order to make the universe behave as you wish, that is to say, give a good account of continents rifting and hummingbirds feeding, it might be that to obtain each additional second of the desired history it is always sufficient to correctly calculate another, say, thousand decimal places for the positions and momenta of the molecules in the initial state.  Or to the contrary, the first thousand decimal places might give you one second, the next thousand only a further half second, then a fourth of a second, and so on.

Yet even if I am wrong in this, we would just go from Scylla to Charybdis.  For in that case classical mechanics must be like the red dust theory, where, from our point of view, anything is possible, or too close to anything.  In either case the classical solution set would imply no structure such as we experience in life.  No sculpted dunes, no ants carting morsels, no shower of hail would pop out of it.  Nor can one imagine any reason why the solution set would show a preference for depicting creatures learning classical mechanics, or if so doing benefiting by it.  In short, there is a total disconnect between the fact that classical mechanics is useful and the hypothesis that the universe as a whole is a classical mechanical system.

That leaves us with an unsolved mystery: why does classical mechanics work for us?  And classical mechanics is the archetype of the kind of physics where we learn from what can be true histories of things.

To my mind, the hourglass with observation of its emitted light is deeply conservative physics.  It makes quantum mechanics as seamless a continuation of the physics of the previous centuries as is at all possible.  This is because of the mathematical form of the hourglass, which is a continuous development from initial conditions, as well as the form of the observations, which impinge as little as can be.  And when this leads to our being given movies rather than direct histories, then I am surprised (and amused) by this, but accept it for the sake of the qualities mentioned, which I consider to be virtues that promise.  Nature is teaching us another lesson.

Bohr\rq s old quantum theory was based on quantum jumps, and I think this was a wonderful piece of exploration in the dark.  When Heisenberg\rq s new quantum mechanics came along, quantum jumps were kept.  The jumps would allow direct histories to be retained as the foundation of our physics, though at the expense of the continuous Hamiltonian evolution of the wave packets (and at the expense of clear definition, for no one has ever been able to specify just when and where and what the quantum jumps are).  Like Schr\"{o}dinger, I~am~jarred by this.  If we are given the choice of preserving philosophical principle or mathematical form, I think we should prefer mathematical form.  Isn\rq t this what Copernicus did?

A final thought:  If learning from the movies provided by hourglasses is how we do physics, then to know why quantum mechanics works would be to know why all the inferences we might make from the movies will fit together with sufficient coherence.  But to know this would require that we know all the things we might ever think of.  It\rq s hopeless.  Though we might nibble at the problem, by showing that the hourglasses have some needed characteristics.  So I think the hourglasses will leave us with an essentially unfathomable mystery.

\referencesz

\noindent Gibbs, J. Willard [1981]: {\it Elementary Principles in Statistical Mechanics\/}, Wood-
\hspace*{\parindent}bridge, CT: Ox Bow Press, p.\hspace{-.5mm} 17 and p.\hspace{-.5mm} 163.

\noindent Mahon, Basil [2003]: {\it The Man Who Changed Everything\/}, Chichester, UK: \hspace*{\parindent}John Wiley \& Sons Ltd.

\vskip4mm

\noindent The hourglasses suggest that von Neumann\rq s measurement theory should be recast for imaginative use rather than for the description of actual situations.  This gives one extra freedom in setting it up, and it can then work more effectively.  An outline is here:

\vskip3mm

\noindent McCartor, Donald [2004]: \lq Quantum Thought Experiments Can Define Na-
\hspace*{\parindent}ture\rq, Concepts of Physics, Vol.\hspace{-1.8mm} I, no.\hspace{-1.8mm} 1, pp.\hspace{-2mm} 105--150 and quant-ph \hspace*{\parindent}0702192.

\vskip5mm

\noindent donaldamccartor@earthlink.net

\end{document}